\documentclass[useAMS,usenatbib]{mn2e}
\usepackage{graphicx}
\usepackage{times}

\def\ltsima{$\; \buildrel < \over \sim \;$}
\def\lsim{\lower.5ex\hbox{\ltsima}}
\def\loe{\lower.5ex\hbox{\ltsima}}
\def\gtsima{$\; \buildrel > \over \sim \;$}
\def\gsim{\lower.5ex\hbox{\gtsima}}
\def\goe{\lower.5ex\hbox{\gtsima}}

\title[Weighing the black holes in ULXs through timing]{Weighing the black holes in ultraluminous X-ray sources through timing}
\author[P. Casella et al.]{P. Casella$^{1}$\thanks{E-mail:casella@science.uva.nl}, G. Ponti$^{2,3,4}$, A. Patruno$^{1}$, T. Belloni$^{5}$, G. Miniutti $^{6}$ and L. Zampieri$^{7}$\\
$^{1}$Astronomical Institute ``Anton Pannekoek'', University of Amsterdam, Kruislaan 403, 1098 SJ Amsterdam, The Netherlands\\
$^{2}$APC Universit\'e Paris 7 Denis Diderot, 75205 Paris Cedex 13, France\\
$^{3}$Dipartimento di Astronomia, Universit\`a degli Studi di Bologna, via Zamboni, 40127 Bologna, Italy\\
$^{4}$INAF-IASF Bologna, via Gobetti 101, 40129 Bologna, Italy\\
$^{5}$INAF-Osservatorio Astronomico di Brera, Via E. Bianchi 46, I-23807 Merate (LC), Italy\\
$^{6}$Institute of Astronomy, University of Cambridge, Madingley Road, Cambridge CB3 0HA, UK\\
$^{7}$INAF-Osservatorio Astronomico di Padova, Vicolo dell'Osservatorio 5, I-35122, Padova, Italy}
\begin{document}

\date{Accepted 2008 April 21. Received 2008 April 11; in original form 2007 December 07.}

\pagerange{\pageref{firstpage}--\pageref{lastpage}} \pubyear{2002}
\maketitle

\label{firstpage}

\begin{abstract}
We describe a new method to estimate the mass of black holes in
Ultraluminous X-ray Sources (ULXs). The method is based on the
recently discovered ``variability plane'', populated by Galactic
stellar-mass black-hole candidates (BHCs) and supermassive active galactic
nuclei (AGNs), in the parameter space defined by the black-hole mass,
accretion rate and characteristic frequency. We apply this method to
the two ULXs from which low-frequency quasi-periodic oscillations have
been discovered, M82 X-1 and NGC 5408 X-1. For both sources we obtain
a black-hole mass in the range $100\sim1300\;M_\odot$, thus providing
evidence for these two sources to host an intermediate-mass
black hole.
\end{abstract}


\section{Introduction} \label{intro}

Ultraluminous X-ray Sources (ULXs) are point-like, off-nuclear sources
which have luminosities greater than $\sim 10^{39}$ erg s$^{-1}$, in
excess of that of a $\sim 10 M_{\odot}$ compact object accreting at
the Eddington limit.
The high luminosity, the very soft thermal component often observed in
many sources (interpreted as emission from a cool accretion disc, see
e.g. \citealt{milleretal03}, and references therein; for different
interpretations of the soft X-ray excess see
e.g. \citealt{stobbartetal06}, and references therein) and the
variability on timescales from months to hours
\citep{swartzetal04,mucciarellietal07} suggest that (some of) these
sources may be powered by accretion onto an Intermediate Mass Black
Hole (IMBH) of 100-1000 $M_{\odot}$. Nevertheless, many of the ULXs
properties can be explained if they do not emit isotropically
\citep[geometrically beamed emission, see][]{kingetal01}, are
dominated by emission from a relativistic jet
\citep[e.g.][]{kaaretetal03}, are accreting at a supercritical rate
\citep[][]{begelman02,ebisawaetal03,begelmanetal06}, or a combination of all
these. In this case, they may harbor stellar mass black holes and may
be similar to Galactic black-hole binaries.

A possible approach to study the nature of ULXs is through fast time
variability. The analysis of the aperiodic variability in the X-ray
flux of X-ray binaries is a powerful tool to study the properties of
the inner regions of the accretion disc around compact objects, since
observed frequencies might be linked to specific time scales in the
accretion disc. The best way to do this is by studying the analogies
of timing features in ULXs with those in stellar-mass black
holes. Measurements of the mass of the black hole in ULXs can then be
derived by applying scaling arguments.

Low-frequency quasi-periodic oscillations (QPOs) have been observed in
the X-ray light curves of two ULXs, M82 X-1
\citep{strohmayermushotzky03} and NGC 5408 X-1
\citep{strohmayeretal07}. The properties of these QPOs (as, e.g.,
fractional amplitude, variability, coherence, amplitude and frequency
of the underlying noise, frequency-flux correlation, etc.; for a
thorough discussion see e.g.
\citealt{mucciarellietal06,strohmayeretal07}) are reminiscent of those
of the most common low-frequency QPO in BHCs (type-C QPO,
\citealt{remillardetal02}; see \citealt{casellaetal05} and references
therein). This association, assuming an inverse scaling of the QPO
frequency with the BH mass, allows for an estimate of the black-hole
masses in the range of a few tens of $M_{\odot}$ to 1000
$M_{\odot}$. In order to further constrain the mass of the black hole
in these two ULXs, several authors explore different possible scaling
relations. \citet{feng&kaaret07} use the correlation between power-law
luminosity and QPO frequency in BHCs to estimate a black-hole mass of
around 10$^3\;M_\odot$ for both ULXs.

From the correlation between the spectral photon index and the QPO
frequency \citep[known to exist in BHCs, see e.g.][]{vignarcaetal03}
\citet{fioritotitarchuk04} derived for the BH in M82 X-1 a mass of the
order of 10$^3\;M_\odot$. From the same correlation,
\citet{strohmayeretal07} derived for the BH mass in NGC 5408 X-1 a
lower limit of 600 $M_{\odot}$. From the correlation between disc
temperature and QPO frequency \citep[known in the BHC XTE
J1550-564,][]{sobczaketal00} the same authors derived a mass in the
range from $\sim$1800 to $\sim$5000 $M_{\odot}$. Other possible QPO
identifications would obviously yield to different mass estimates
\citep[although these identifications are more unlikely, for a
thorough discussion see][]{mucciarellietal06,strohmayeretal07}.

All the mass measurements/constraints mentioned above make use of
correlations known to exist in stellar-mass black holes. This approach
is powerful and physically intriguing, although it requires a
non-trivial assumption: that the slope of these correlations, derived
for the stellar-mass BHCs, remains the same over wide ranges of
frequencies, luminosities and - possibly - masses. In this Letter we
explore a new method to estimate the mass of black holes in ULXs. This
method is based on the correlation between characteristic frequencies
\citep{bpk02}, on the ``fundamental plane'' \citep{merlonietal03} and on
the ``variability plane'' \citep{kordingetal07} of accreting black
holes.  The first correlation is known to hold for neutron stars and
stellar-mass black holes, hence an assumption is required to extend it
to the ULXs (see \S\ref{sec:freq}). On the other hand, the fundamental plane and
the variability planes are populated by accreting black holes over a
wide range of masses, from the stellar-mass BHCs to the supermassive
black holes in active galactic nuclei (AGNs).

\begin{figure}
\begin{center}
\includegraphics[width=7.0cm]{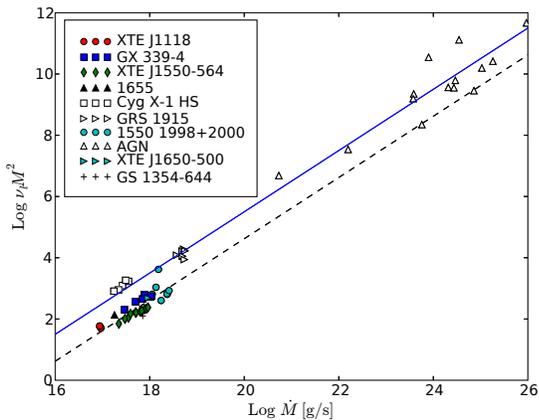}
\end{center}
\caption{Variability plane for stellar-mass BHCs and AGNs. The lines
indicate $\nu\propto \dot{M}M^{-2}$. The upper line is a fit to
soft-state objects, the lower is a fit to the hard-state BHCs only
\citep[from][]{kordingetal07}.}
\label{kording}
\end{figure}

\section{The variability plane} \label{method}

\citet{mchardyetal06} has shown that soft-state BHCs and active
galactic nuclei (AGNs) populate a plane in the parameter space defined
by the black-hole mass, accretion rate and characteristic frequency in
the PDS. \citet{kordingetal07} recently showed evidence for this
variability plane to extend to the hard-states BHCs, with a
constant offset for the frequencies (see Fig.~\ref{kording}):

\begin{equation} \label{eq:kording}
\log\nu_\ell = \log\dot{M} - 2 \log M - 14.7 - 0.9\;\theta
\end{equation}

\noindent where $\theta$ goes from 0 for the soft state to 1 for the
hard state (from eq.~6 of \citet{kordingetal07}). The existence of
such a variability plane across different spectral states, if
confirmed, would suggest that accretion on BHs is scale invariant. If
this is the case, we can thus use the correlation to measure the mass
of the black holes thought to be hosted in ULXs. The frequencies and
the luminosities observed in ULXs are somewhat intermediate between
those in BHCs and in AGNs. Hence it is reasonable and relatively
straightforward to put them in the variability plane. The strength of
this method is that it uses a relation which appears to hold for many
orders of magnitude in frequency, luminosity and black-hole
mass. Either ULXs follow the same relation of all the accreting black
holes, thus yielding constraints on the masses of the black holes they
host, or they do not, thus demonstrating that their accretion flow is
somewhat different than in BHCs and AGNs.

\begin{figure}
\begin{center}
\includegraphics[width=7.1cm]{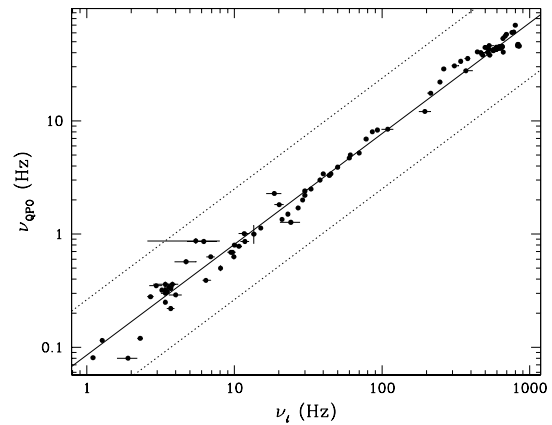}
\end{center}
\caption{The QPO centroid frequency ($\nu_{QPO}$, which corresponds to
the QPO observed in M82 X-1 and NGC 5408) versus the lower
high-frequency Lorentzian \citep[$\nu_\ell$, which is the one used for
the variability plane in][]{kordingetal07} in neutron stars and BHCs
\citep[adapted from][]{bpk02}. The solid line shows the power-law fitted to the data (Eq.~\ref{eq:freq}), while the two dotted lines show the correspondent
uncertainty of $\pm0.5$~dex we adopt in our estimate of $\nu_\ell$ (see
Sect.~\ref{sec:uncert}).}
\label{bpk}
\end{figure}

\subsection{Estimate of the characteristic frequency} \label{sec:freq}

In order to use the variability plane to measure the mass of a
black hole, we must correctly identify the characteristic frequencies
in the PDS of the X-ray light curve of the black hole itself. The
relatively robust identification of the QPOs in M82 X-1 and NGC 5408
X-1 with the ``type-C'' QPO in BHCs (see Sec.~\S\ref{intro}), helps us
in this. The frequency of the type-C QPO is known to correlate with
other characteristic frequencies measured in the PDS of BHCs
\citep[][and references
therein]{wvdk99,pbk99,bpk02,kleinwolt&vdk07}. In particular, the
frequency of the component used in \citet{kordingetal07} (the lower
high-frequency Lorentzian $\nu_\ell$) is found to be a factor of
$\sim$10 higher than the centroid frequency of the type-C QPO (see
Fig.~\ref{bpk} and Eq.~\ref{eq:freq}). In Figure~(\ref{bpk}) we report
data from \citet{bpk02} and the best fit power-law to them, after
removal of points which do not belong to the plot (because of a
different and/or dubious identification of the two frequencies, see
\S~4.3 of \citet{bpk02}).

If we assume that this correlation holds also for
ULXs, we can use it to estimate $\nu_\ell$ for M82 X-1 and NGC 5408
X-1 from the QPO frequency $\nu_{QPO}$ :

\begin{equation} \label{eq:freq}
\nu_\ell \approx 12.37 \times (\nu_{QPO})^{1.023}
\end{equation}

\noindent It must be noted however, that this relation is obtained by
extending to the ULXs the relation in Fig.(\ref{bpk}), which is known
to hold only for stellar-mass compact objects. This extensions appears
to us reasonable, given the fact that the relation appears rather tight
over a very broad range of frequencies (which includes the frequencies
of the QPOs observed in M82 X-1) and compact-object masses, but it is
not certain and will need to be confirmed by the next generation of
X-ray satellites, which will be able to detect the high-frequency
component $\nu_\ell$ in ULXs.

\subsection{Estimate of the accretion rate} \label{sec:mdot}

Another key point is of course how to estimate the accretion rate from
the X-ray luminosity. To do this, we need to know how efficient is the
accretion in ULXs. Many observational evidences suggest that M82 X-1
and NGC 5408 X-1 are in a somewhat intermediate state \citep[see
e.g.][]{mucciarellietal06}. The main indication for this is the high
amplitude of the QPOs observed in these sources, since BHCs do not
show strong QPOs in their soft state \citep[see
e.g.][]{bellonietal339}. This means that, a priori, the accretion can
be either efficient (soft-state like) or inefficient (hard-state
like), or, most probably, intermediate between the two. We can
calculate the accretion rate under the two extreme hypotheses, thus
deriving a range for the mass of the black hole.

\subsubsection{Efficient case} In case ULXs are efficiently
accreting sources ($\eta\sim0.1$), we can measure the accretion rate
directly from the bolometric luminosity:

\begin{equation} \label{eq:soft:mdot}
\dot{M}={{L_{bol}}\over{0.1 c^2}}>{{L_{X}}\over{0.1 c^2}}.
\end{equation}

\noindent Where the lower limit is due to the fact that we don't know
the conversion factor from $L_X$ to $L_{bol}$. Applying
Eq.~(\ref{eq:freq}) and (\ref{eq:soft:mdot}) to the
Eq.~(\ref{eq:kording}) for the soft-state case ($\theta=0$), we can
thus derive an expression for the mass of the black hole:

\begin{equation} \label{eq:soft:mass}
\log\left(\frac{M}{M_{\odot}}\right) > \frac{\mbox{1}}{\mbox{2}}\log \left(\frac{L_X}{0.1 c^2}\right) - \frac{\mbox{1}}{\mbox{1.95}}\log \nu_{QPO} - 7.9
\end{equation}

\subsubsection{Inefficient case} If the accretion is
inefficient, the accretion rate is not linearly related to the X-ray
luminosity. However, inefficiently accreting black holes are known to
lie on the so-called ``fundamental plane''
\citep{merlonietal03,falckeetal04} defined by X-ray luminosity, radio
luminosity and mass. This relation holds for many order of magnitudes
in masses, connecting the Galactic stellar-mass BHCs to the
extragalactic supermassive black holes in AGNs. Under the reasonable
assumption that inefficiently-accreting BHs in ULXs would also lie on
the fundamental plane (see the Discussion), we can estimate the
accretion rate from the 2-10 keV X-ray luminosity:

\begin{equation} \label{eq:hard:mdot}
\dot{M} \approx 0.12\;(L_X)^{0.5} M^{0.43}\;\left(\frac{\mbox{0.1}}{\eta_{accr}}\right)\;\mbox{g s}^{-1}
\end{equation}

\noindent where $\eta_{accr}$ is the total efficiency of the accretion flow,
including both radiative emission and kinetic energy of any
ejected matter \citep[from eq.~8 of][]{kordingetal06}.

Using again Eq.~(\ref{eq:kording}), this time for
the hard-state case ($\theta=1$), and replacing $\nu_\ell$ and
$\dot{M}$ with our Eqs.~\ref{eq:freq} and \ref{eq:hard:mdot}, we
derive the expression for the mass of the black hole for the
inefficient-accretion case:

\begin{equation} \label{eq:hard:mass}
\log M = \frac{\mbox{1}}{\mbox{3.14}}\log L_X - \frac{\mbox{1}}{\mbox{1.53}}\log \nu_{QPO} - 11.226 .
\end{equation}

\noindent Eq.~(\ref{eq:soft:mass}) and (\ref{eq:hard:mass}) give the
bounds for the mass values for a black hole in a ULX with an X-ray
luminosity $L_{X}$ and a characteristic frequency $\nu_{\ell}$ in the
PDS.


%
%
%
%

\subsection{Uncertainties} \label{sec:uncert}

The uncertainties on the mass values given from
Eq.~(\ref{eq:soft:mass}) and (\ref{eq:hard:mass})
mainly come from the intrinsic scatter of the three correlations we
used to derive the mass estimates. These scatters are larger than the
observational uncertainties in the measurements of fluxes and QPO
frequencies.

From Fig.~\ref{bpk}, we see that the full scatter in $\nu_{\ell}$ is
always smaller than half a decade. We can thus associate a very
conservative uncertainty of 0.5 dex (i.e. 1 dex scatter) to the values
of $\nu_{\ell}$ obtained from the measurement of $\nu_{QPO}$
(Eq.~\S\ref{eq:freq}). In other words, given a measurement of
$log(\nu_{QPO})$, the derived value of $log(\nu_{\ell})$ will have a
symmetric uncertainty $\Delta_{\nu} = 0.5$. This uncertainty obviously
dominates the one on the direct measurement of $\nu_{QPO}$ (which is
of the order of $\sim 1\%$).

The uncertainty on the accretion rate estimate is dominated from the
scatter in the fundamental plane which, to be conservative, we estimate
to be of the order of $\sim$0.5~dex \citep[see fig. 4
of][]{kordingetal06}, which translates in a symmetric uncertainty $\Delta_F = 0.25$ dex.

Since we use the variability plane to estimate the black-hole masses,
the intrinsic scatter of the plane itself gives of course an important
contribution to the uncertainty on the mass
values. \citet{kordingetal07} extensively discuss the uncertainties of
the variability plane, and conclude estimating an overall error on the
normalization of 0.22 dex. The dependent variable in the variability
plane is $\nu_{\ell} M^2$, which translates in a scatter (full range)
of 0.11 dex on the mass (i.e., a symmetric uncertainty $\Delta_K =
0.055$ dex). An uncertainty of $\sim 10 \%$ ($\Delta_L = 0.1$) can be
associated to the measurement of the X-ray luminosity.

These uncertainties are clearly independent, hence we can propagate
them quadratically through the Eq.~(\ref{eq:soft:mass}) and
(\ref{eq:hard:mass}). We do not associate any error to the estimate of
the accretion rate for the efficient case. As we do not convert the
X-ray luminosity in bolometric luminosity, we are considering a
lower limit for $\dot{M}$.

The total uncertainties $\Delta_M$ on the values of $log(M)$ are thus:

\begin{equation} \label{eq:soft:err}
\Delta_{M} \approx \sqrt{\left(\frac{\Delta_L}{2}\right)^2 + \left(\frac{\Delta_{\nu}}{1.95}\right)^2 + \Delta_K^2 + \Delta_F^2} = 0.36
\end{equation}

\noindent for the efficient case (Eq.~\ref{eq:soft:mass}), and:

\begin{equation} \label{eq:hard:err}
\Delta_{M} \approx \sqrt{\left(\frac{\Delta_L}{3.14}\right)^2 + \left(\frac{\Delta_{\nu}}{1.53}\right)^2 + \Delta_K^2 + \Delta_F^2} = 0.41
\end{equation}

\noindent for the inefficient case (Eq.~\ref{eq:hard:mass}).


\section{The cases of M82 X-1 and NGC 5408 X-1} \label{sec:sources}

Let us now apply these arguments to the two ULXs for which a QPO has
been discovered: M82 X-1 and NGC 5408 X-1.

\subsection{M82 X-1} \label{sec:sources:m82}

The QPO in M82 X-1 was observed at 54 mHz and 112 mHz in two {\it
XMM-Newton} observations in 2001 \citep{strohmayermushotzky03} and
2004 \citep{mucciarellietal06,dewanganetal06}. From these frequencies
we derive a $\nu_\ell$ of 0.54 Hz and 1.12 Hz, respectively (see
Sec.\S\ref{sec:freq}). After correcting for crowding,
\citet{feng&kaaret07} estimate a 2-10 keV source luminosity of $1.3
\times 10^{40} \mbox{ergs s}^{-1}$ and $1.7 \times 10^{40} \mbox{ergs
s}^{-1}$ for the 2001 and 2004 observation, respectively. By inserting
these numbers in Eq.~(\ref{eq:soft:mass}) we obtain, for the efficient
case, a black-hole mass $700_{-395}^{+905}\;M_\odot$ (2001) and
$550_{-310}^{+710}\;M_\odot$ (2004). If we assume an inefficient accretion, we
instead obtain a black-hole mass value of $243_{-148}^{+380}\;M_\odot$ (2001)
and $165_{-100}^{+260}\;M_\odot$ (2004) from Eq.~(\ref{eq:hard:mass}).

From these values we see that we can put a lower limit for the black
hole in M82 X-1 of $95\;M_\odot$ (smaller values would not be
consistent with the 2001 observation) and an upper limit of
$1260\;M_\odot$ (higher values would not be consistent with the 2004
observation).

\subsection{NGC 5408 X-1} \label{sec:sources:ngc5408}

The QPO in NGC 5408 X-1 was observed at 20 mHz
\citep{strohmayeretal07}, which yields to a $\nu_\ell=0.20$~Hz. We
re-analyzed the public {\it XMM-Newton} data and obtained a 2-10 keV
unabsorbed luminosity of $3.0 \times 10^{39} \mbox{ergs
s}^{-1}$. Using these values, we derive a black-hole mass of
$295_{-180}^{+465}\;M_\odot$ for the inefficient-accretion case and
$570_{-320}^{+735}\;M_\odot$ for the efficient-accretion case. This
translates in a lower limit of $115 M_\odot$ and an upper limit of
$1300 M_\odot$ for the mass of the black hole in this source.

\begin{table}
 \centering
 \begin{minipage}{83mm}
  \caption{Values of the characteristic frequencies, luminosities and
  inferred masses for M82 X-1 and NGC 5408 X-1.}
 \label{values}
  \begin{tabular}{@{}lccccc@{}}
  \hline
   Source             & $\nu_{QPO}$  & $L_{X}$ \footnote{$\times\;10^{38}$, in the 2-10 keV range.} & \multicolumn{2}{c}{Black-hole mass ($M_{\odot}$)} \\
                      &     (mHz)   &        (ergs/s)                      &    Ineff. accr.   &   Eff. accr.          \\
 \hline
  M82 X-1~(2001) & 54$\pm$1   & 130$\pm$13 & 240$_{-150}^{+380}$ & 700$_{-395}^{+905}$ \\

  M82 X-1~(2004) & 113$\pm$2  & 170$\pm$17 & 165$_{-100}^{+260}$  & 550$_{-310}^{+710}$ \\

  NGC~5408 X-1   & 20$\pm$0.5 &  30$\pm$3 & 295$_{-180}^{+465}$ & 570$_{-320}^{+735}$ \\
  \hline
\end{tabular}
\end{minipage}
\end{table}

\section[]{Discussion} \label{discussion}

In Table \ref{values} we report the inferred values for the mass of
the black holes in M82 X-1 and NGC 5408 X-1. It is evident that this
method supports the identification of both black holes as
``intermediate-mass black holes''. Since we do not know the efficiency
of the accretion in the two sources, we cannot further constrain the
masses. However, a comparison with BHCs can give some hints on this.

QPOs similar to those discovered in these two ULXs are often observed
in BHCs during their intermediate state. The frequency of these QPOs
usually increases, and their amplitude decreases, as the source
becomes brighter and the energy spectrum softens (see
e.g. \citealt{casellaetal04,bellonietal339,homanbelloni05}). This
softening is usually due to a steepening of the hard power law and to
an increase of the soft, thermal disc flux. The increase of the disc
flux might be also the origin of the decrease of the QPO amplitude, in
case the QPO itself arises from the hard power law.  Hence, the high
amplitude of the QPOs observed in the X-ray light curve of M82 X-1 and
NGC 5408 X-1, as well as the presence of a very weak disc in their
energy spectra, suggest that these two sources are on the hard side of
the intermediate state. However, in case these two sources host an
intermediate-mass black hole, a relatively low-temperature disc is
expected \citep[][and references therein]{milleretal03}.

The QPO in M82 X-1 has already been observed at two different
frequencies. Over this small data set the source has shown to follow
several correlations known to exist in BHCs. Namely, the QPO frequency
increases and its amplitude decreases as the source becomes
brighter. This means that during the 2004 observation (see Table
\ref{values}) the source was probably in a slightly softer, more
efficient state than in 2001. If we use, for the 2004 observation, the
same luminosity-to-accretion rate conversion than in 2001, we are
actually underestimating the accretion rate. This is consistent with
the fact that we obtain slightly lower ranges of mass values for the
2004 observation than for the 2001 one.

In Section \S\ref{sec:mdot} we use the fundamental plane of accreting
BHs to convert the X-ray luminosity in accretion rate for the
inefficient case. This method is based on the assumption that ULXs lie
on the same radio/X-ray correlation as BHCs and AGN do. To date, radio
counterparts have been found only for a few ULXs \citep[see e.g.][and
references therein]{kordingetal05}. A full discussion about the
radio-to-X-ray luminosity ratio of ULXs has already been started by
other authors, and is beyond the aim of this Letter. Here we only note
that NGC 5408 X-1 is one of the few ULXs for which a steady radio
counterpart has been detected. \citet{kordingetal05} used the values
of X-ray and radio luminosity of this source to measure the mass of
the hosted black hole, obtaining a value of $\sim 1000\;M_\odot$,
which is roughly consistent with the values we found. It must be
noted, however, that the observed steep, inverted radio spectrum is
consistent with a thin synchrotron emission, and not with the steady
flat radio spectrum observed in hard-state BHCs \citep[for a
discussion see][]{soriaetal06}. The slope of the radio emission in NGC
5408 X-1 suggests that the source is on the soft side of the
intermediate state, when thin synchrotron radio emission is often
observed in BHCs. This would imply that the mass of the BH in this
source is closer to the values obtained in the efficient case.  A
bright, variable radio counterpart has been reported also for M82
X-1. Its strong variability clearly rules out the possibility to use
the radio flux to place the source on the fundamental plane, since the
latter is valid only for steady states. The upper limit for the steady
emission, however, is still consistent with the source lying on the
plane and a BH of a few hundreds of solar masses.

\subsection[]{Non-standard accretion flow} \label{disc:nonstandard}

The method described in this Letter is based on the initial assumption
that the accretion flow in ULXs follows the same general rules as the
one in stellar-mass and supermassive BHs does. If the accretion onto
the black holes in ULXs does not follow general scale-invariant
relations, the method described here, as well as many other scaling
arguments, loses validity. Let us thus discuss possible alternative
scenarios. 

For example, it has been proposed for the ULXs to be sources similar
to SS433, but with a different viewing angle. In this scenario, most
of the X-ray luminosity emitted from a supercritical accretion disc
\citep{begelman02} would be geometrically collimated
\citep{begelmanetal06,poutanenetal07}. The main concern, when invoking
a beamed emission (either geometric or relativistic) is whether QPOs
can be preserved. Any physical interpretation of the ULXs involving
beaming will need to demonstrate that relatively coherent oscillations
in the X-ray flux do not get smeared out by the beaming itself.

Without applying any beaming correction, i.e. remaining under the
assumption of roughly isotropic emission, there are only two possible
ways for a stellar-mass BH ($\lsim 20\;M_\odot$) to reach X-ray
luminosities as high as $10^{40} \;\mbox{ergs s}^{-1}$: either the
accretion rate is super-Eddington, or the accretion rate is still
below the Eddington limit, but the accretion flow is extremely
efficient, as to produce such high luminosities. For example, it has
been suggested that ULXs host super-Eddington accreting, stellar-mass
black holes, somewhat similar to the microquasars in our Galaxy, as
GRS 1915+105 \citep[see e.g.][]{king02}. However, GRS 1915+105 itself
lies on many of the relations used here and in literature to measure
the black-hole mass in ULXs, as well as other very luminous
microquasars do. The very highly, sometimes super-Eddington accreting
Galactic microquasars appears to follow the same general relations of
the sub-Eddington accreting black holes.

To discuss the case of extremely efficient accretion, let us take the
most extreme case of an efficiency of 50 \% (maximally rotating Kerr
BH) {\it and} of a bolometric luminosity equal to the X-ray
luminosity. Under these extreme assumptions, from
Eq.~(\ref{eq:soft:mass}) we would obtain
M~=~300$_{-170}^{+400}\;M_{\odot}$ (for the 2004 observation of M82
X-1). This demonstrate that, even in case of an extremely efficient
accretion, the method presented in this Letter yields a BH mass 
higher than $130\;M_{\odot}$.

More observations are clearly needed in order to test the validity of
the method described here. Detecting more QPOs from M82 X-1 and NGC
5408 X-1, hopefully at different frequencies and/or luminosities, will
show whether these two sources lie on the variability plane or not.
The current data strongly argue for M82 X-1 and NGC 5408 X-1 to host
intermediate-mass black holes, both with a mass from $\sim 100$ to
$\sim 1300~M_\odot$.

%

\section*{Acknowledgments}

PC would like to thank Rudy Wijnands for very useful discussions and
comments on the paper. We thank the anonymous referee for his/her
careful reading and very useful comments. TB acknowledges support from
PRIN-INAF 2006 grant.  GP acknowledges support from ASI/INAF contract
number I/023/05/0. This work was partially supported by the
Netherlands Organization for Scientific Research (NWO) and by the
Netherlands Research School for Astronomy (NOVA).

\bsp

\label{lastpage}

\end{document}